\documentclass[12pt]{iopart}

\usepackage[english]{babel}
\usepackage{latexsym}

\usepackage{color}
\usepackage{graphics}
\usepackage{subfigure}
\usepackage{epsfig}

\usepackage{setstack}
\usepackage{iopams}
\usepackage{bbm}
\newcommand{\bra}[1]{\ensuremath{\langle#1|}}

\newcommand{\ket}[1]{\ensuremath{|#1\rangle}}

\newcommand{\braket}[2]{\ensuremath{\langle #1|#2\rangle}}

\newcommand{\eins}{\ensuremath{\mathbbm 1}}

\newcommand{\BE}{\begin{equation}}
\newcommand{\EE}{\end{equation}}
\newcommand{\be}{\begin{equation}}
\newcommand{\ee}{\end{equation}}
\newcommand{\bea}{\begin{eqnarray}}
\newcommand{\eea}{\end{eqnarray}}
\newcommand{\kommentar}[1]{}

\newcommand{\ew}[1]{\ensuremath{\langle #1 \rangle}}

\newcommand{\ie}{{\it i.e.}}

\newcommand{\via}{{\it via\ }}

\newcommand{\beq}{\begin{equation}}
\newcommand{\eeq}{\end{equation}}
\newcommand{\beqa}{\begin{eqnarray}}
\newcommand{\eeqa}{\end{eqnarray}}
\newcommand{\bse}{\begin{subequations}}
\newcommand{\ese}{\end{subequations}}

\newcommand{\eqnref}[1]{Eq.~(\ref{#1})}

\newcommand{\figref}[1]{Fig.~\ref{#1}}

\newcommand{\vecb}[1]{\hat{\mbox{\boldmath$#1$}}}

\begin{document}

\title{Efficient quantum state transfer in spin chains via adiabatic passage}

% For NJP
\author{K. Eckert$^1$, O. Romero-Isart$^1$, and A. Sanpera$^{1,2}$}
\address{$^1$ Departament F\' isica, Grup F\' isica Te\'orica, Universitat Aut\'onoma de Barcelona, E-08193 Bellaterra, Spain.}
\address{$^2$ ICREA: Instituci\'o Catalana de Recerca i Estudis Avan\c cats.} 
\ead{kai@uab.es}

\date{\today}

\begin{abstract}
We propose a method for quantum state transfer in spin chains using an adiabatic passage technique.
Modifying even and odd nearest-neighbour couplings in time allows to achieve transfer fidelities arbitrarily close
to one, without the need for a precise control of coupling strengths and timing.
We study in detail transfer by adiabatic passage in a spin-1 chain governed by
a generalized Heisenberg Hamiltonian. We consider optimization
of the transfer process applying optimal control techniques. We discuss a realistic experimental
implementation using cold atomic gases confined in deep optical lattices.
\end{abstract}
%\end{frontmatter}

\maketitle

\section{Introduction}

Reliable quantum communication between distant nodes in a quantum network
is of uttermost importance for quantum computation and
information \cite{bennett:2000}. For relatively short distances, since the
pioneering works of S. Bose \cite{bose:2003:b} and V. Subrahmanyam \cite{subrahmanyam:2004}
spin chains are considered to be good candidates to perform this task.  However, due to dispersion
in the free evolution, the fidelity of the transmission through such a chain decreases as the number of spins is increased. 
This problem was first circumvented through local engineering of the (nearest-neighbor) couplings in the chain to
obtain a perfect channel \cite{christandl:2004}.  Since then, many schemes
have been proposed to maximize transfer fidelity and practicability. Coupling the outer spins only weakly
to the chain allows to perturbatively obtain an effective Hamiltonian for the two outer sites only
\cite{weak1,weak2}. Also, dispersion can be minimized by suitably encoding a quantum state in several neighbouring spins \cite{gauss,hasel:2006}.
If the quantum state is encoded in two parallel, uncoupled spin chains ('dual-rail' encoding),
then a suitable measurement at the end of the chain allows to conclude that the state has been transferred
successfully \cite{dual}. All these schemes rely on chains or ladders of spin-$1/2$s. Transfer of gaussian states has been studied in
chains of harmonic oscillators \cite{harmonic1,harmonic2}. For most protocols, no temporal control of the
couplings inside the chain is necessary. However, usually it is inevitable to control the first and the
last coupling in order to start the transport and to receive the state at the other side, respectively.
Especially the control of the last coupling often is crucial to retrieve the state with maximal fidelity.

Here we study quantum state transfer in one-dimensional (1D) spin-$1$ systems, requiring a (limited)
temporal control of all the couplings in the chain. We focus on the most general isotropic
spin-$1$ Hamiltonian, the generalized Heisenberg model.
Its rich phase diagram, which includes several paradigmatic anti--ferromagnetic phases of strongly correlated
many--body systems, has been intensively studied for the last two decades (see, e.g., \cite{legeza} and references therein).
Still, properties of ground states and low excitations are under debate in certain regions \cite{legeza}.
Recently, this type of Hamiltonians has attracted a renewed interest because their
clean realization is possible by loading spin-1 atoms (e.g., ${}^{87}$Rb or
${}^{23}$Na) into a deep 1D optical lattice \cite{imambekov:2003,garcia-ripoll:2004,rizzi:2005}.
Quantum state transfer through spin-1 chains governed by the generalized Heisenberg Hamiltonian
has been studied in \cite{romero-isart:2006} for the case when the chain with all couplings
fixed and identical is prepared in its ground state. Like in the spin-$1/2$ case, the free
evolution after a single spin is coupled to the chain presents dispersion. Therefore the transfer is generically
non-perfect. However, certain (anti-ferromagnetic) regions in the phase diagram give rise to particularly high transport
fidelities. The transfer mechanisms in these cases are strikingly different from those in spin-$1/2$ chains which
usually start from a ferromagnetic state.

Here we propose a scheme that uses adiabatic passage techniques to obtain arbitrarily perfect state transfer
\footnote{
During completion of this manuscript, an article of T. Ohshima {\it et al.} has appeared \cite{ohshima:2007} which
discusses quantum state transfer in spin-$1/2$ chains through adiabatic dark state passage.
}.
It requires the ability to modify even and odd couplings in the chain independently.
Such a control can be  realized for atoms confined in an optical lattice \via
super-lattice techniques. By choosing the duration of the process long enough, the scheme
allows to increase the transport fidelity arbitrarily close to unity, without the need
for a precise control of the couplings.
A transfer scheme using adiabatic passage techniques is inevitably slow as compared to the speed of propagation
of a low excitation in the system with all couplings active and identical (the same happens for schemes using local
engineering of couplings or working in the perturbative regime). We will, however, demonstrate that optimal control
techniques allow to significantly reduce the duration of the transport process.

The paper is organized as follows: In Section \ref{sec:bbh} we introduce the
spin-1 generalized Heisenberg Hamiltonian with bilinear-biquadratic nearest-neighbor couplings. In
Section \ref{sec:transfer} we provide a description of the transfer scheme and
discuss in detail its performance. In Section \ref{sec:fid} we consider optimization
\via optimal control techniques. A realistic implementation
of the proposed scheme in a system of ultracold atoms confined
in a deep optical lattice is discussed in Section \ref{sec:impl}.
Finally, we conclude in Section \ref{sec:concl}.

\section{Generalized Heisenberg Hamiltonian}\label{sec:bbh}

We consider a one-dimensional chain of $N$ spin-1s, coupled through
the most general isotropic 1D Hamiltonian (the generalized Heisenberg, or bilinear-biquadratic, Hamiltonian)
with nearest-neighbor interactions
\begin{eqnarray}
\hat H_{\rm bb}&=&\alpha\sum_{i}\left[\cos\theta\left(\vecb{J}_i\vecb{J}_{i+1}\right)+
\sin\theta\left(\vecb{J}_i\vecb{J}_{i+1}\right)^2\right]\label{eq:H1}\\
&\equiv&\alpha\sum_i\hat H^{(i,i+1)}_{\rm bb},\nonumber
\end{eqnarray}
Here, $\vecb{J}_i$ denotes the vector of spin-1 operators for site $i$ and $\alpha\geq0$ fixes the
coupling strength. The first term in $\hat H_{\rm bb}$
is the usual linear Heisenberg interaction. The second, quadratic, Heisenberg term does not appear
in spin-1/2 systems, as in that 
case any power of $(\vecb{J}_i\vecb{J}_{i+1})$ can be
expressed as a combination of a constant and a linear term. 
The properties of $\hat H_{\rm bb}$ often become more intuitive if it is expressed in terms of two-site
operators $P^{(i,i+1)}_{S}$, projecting sites $i,\,i+1$ onto manifolds with total spin $S$.
Using the relation
\beq
\left(\vecb{J}_i\vecb{J}_{i+1}\right)^n=\sum_{S=0}^2 \left(\frac{S(S+1)}2-2\right)^n P^{(i,i+1)}_{S},
\eeq
we obtain
\beq
\hat H_{\rm bb}^{(i,i+1)}=\lambda_0 \hat P^{(i,i+1)}_0
+\lambda_1 \hat P^{(i,i+1)}_1+\lambda_2 \hat P^{(i,i+1)}_2.\label{eq:H1b}
\eeq
%\begin{enarray}
%P^{(i,i+1)}_0=-\frac{\eins}3+\frac13\left(\vecb{J}_i\vecb{J}_{i+1}\right)^2,\\
%P^{(i,i+1)}_1=\eins-\frac12\vecb{J}_i\vecb{J}_{i+1}-\frac12\left(\vecb{J}_i\vecb{J}_{i+1}\right)^2,\\
%P^{(i,i+1)}_2=\frac{\eins}3+\frac12\vecb{J}_i\vecb{J}_{i+1}+\frac16\left(\vecb{J}_i\vecb{J}_{i+1}\right)^2,
%\end{eqnarray}
with $\lambda_0=-2\cos\theta+4\sin\theta,\,\lambda_1=-\cos\theta+\sin\theta,$
and $\lambda_2=\cos\theta+\sin\theta$ \cite{imambekov:2003}. Note that is always possible to 
set the smallest $\lambda_{i}$ to zero by adding a multiple of the identity to $\hat H_{\rm bb}$.

Properties of this spin-1 Hamiltonian, \eqnref{eq:H1}, especially of its ground-states, have been
extensively studied (see, e.g, \cite{legeza,imambekov:2003,rizzi:2005,aklt,yip:2003,trebst:2006}. It is the interplay
between the bilinear and biquadratic coupling which gives rise to a rich manifold of ground states.
The ground state is ferromagnetic (\ie, magnetized) for $\frac12\pi<\theta<\frac54\pi$ and anti-ferromagnetic (\ie, non-magnetized) otherwise.
Here we concentrate on the fully dimerized phase, $\theta=-\pi/2$. To better understand its properties,
let us start by considering a system of only two coupled sites.
Since for $\theta=-\pi/2$ the two-site Hamiltonian takes (after a constant shift)
the simple form $\hat H_{\rm bb}^{(i,i+1)}=-\hat P_0^{(i,i+1)}$, the energy is minimized by a singlet state
$\ket{s}_{12}=\left(\ket{1}_1\ket{-1}_2+\ket{-1}_1\ket{1}_2-\ket{0}_1\ket{0}_2\right)/\sqrt3$.
Here we denote the three eigenstates of $\hat J_z$ as $\ket{\pm 1}$, $\ket{0}$.
For more than two sites, such a configuration cannot be repeated on neighboring bonds. Still,
for an even number of sites and open boundary conditions, a good caricature of the ground state is
given by a dimer (or valence bond) state, which has a singlet on each second bond.
Though we concentrate here on $\theta=-\pi/2$, the scheme we propose
works in a region of the phase diagram around this point.

\section{Transfer scheme}\label{sec:transfer}

To achieve the state transfer, we assume an odd number of sites, $N=2n+1$,
and modify the Hamiltonian \eqnref{eq:H1} as follows (here and in the following we set $\theta=-\pi/2$):
\beq
\hat H_{\rm bb}(x)=\alpha\sum_{i=1}^{N-1}\frac{1+(-1)^ix}2\,\hat H_{\rm bb}^{(i,i+1)},\label{eq:H2}
\eeq
{\it i.e.}, we introduce the parameter $x\in[-1,1]$ to be able to adjust the ratio between even and odd couplings.
For $x=1$ ($x=-1$) only even (odd) couplings are non-zero.
Realization of such a Hamiltonian for ultracold atoms (as ${}^{23}$Na) confined in an optical lattice
is discussed in Sec.~\ref{sec:impl}. 

Let us start by analyzing in detail a system of $N=3$ sites. % and $\theta=-\pi/2$.
%Then, except for a constant, we have
%$\hat H_{\rm bb}(x)=-\alpha\sum_i(1+(-1)^ix)\hat P_0^{(i,i+1)}/2$.
Setting initially $x=1$, only the coupling between sites $2$ and $3$ is turned on, while the first site is decoupled.
The (degenerate) ground state then consists of an arbitrarily oriented spin at site $1$ and the two other spins
paired into a singlet. Thus
\beq
\ket{\psi_{x=1}^0(\phi)}=\ket{\phi}_1\ket{s}_{23}.
\eeq
On the other hand, if $x=-1$, only the first two spins are coupled.
In the ground state they then form a singlet, while the last spin is in an arbitrary state $\ket{\phi'}$:
\beq
\ket{\psi_{x=-1}^0(\phi')}=\ket{s}_{12}\ket{\phi'}_3.
\eeq
We will now show that $\ket{\psi_{x=1}^0(\phi)}$ and $\ket{\psi_{x=-1}^0(\phi)}$ are connected
via an adiabatic path as $x$ is changed continuously from $1$ to $-1$. 
Then, a protocol to transfer an arbitrary state $\ket{\phi}$ between
the two spins at the end of the chain can be constructed as follows:
(i) at $t=t_{\rm start}$ choose $x=1$ and prepare the system in its ground state, with the first spin initialized
to the state $\ket{\phi}$ to be transported:
$\ket{\psi(t=t_{\rm start})}=\ket{\psi_{x=1}^0(\phi)}$; (ii) change 
$x$ to reach $x=-1$ at $t=t_{\rm final}$. If the total duration of the process
$T=t_{\rm final}-t_{\rm start}$ is long enough to have a sufficiently slow change of
$x$, then according to the adiabatic theorem \cite{b:schiff} the system remains in the same band of instantaneous eigenstates,
and $\ket{\psi(t=t_{\rm final})}\approx\ket{\psi_{x=-1}^0(\phi)}$. Note that a precise
timing for decoupling and/or reading out the final spin after the transport is not crucial here
(as it is for transport schemes using fixed couplings \cite{bose:2003:b,christandl:2004,subrahmanyam:2004}),
as the coupling of the last spin is adiabatically set to zero.

We take $\ket{\phi}=\ket{\phi'}=\ket{1}$ and denote $\ket{L}\equiv\ket{1}_1\ket{s}_{23},\,\ket{R}\equiv\ket{s}_{12}\ket{1}_3$.
Observing $P_0^{(12)}\ket{L}=\ket{R}/3$,
we find that $\hat H_{\rm bb}(x)$ only acts in the subspace spanned by
$\{\ket{L},\,\ket{R}\}$.
As $\braket{L}{R}=1/3$, we introduce an orthonormal basis as
\beq
\big\{\ket{b_1}\equiv\frac{\sqrt3}{2\sqrt2}(\ket{L}+\ket{R}),\,\ket{b_2}\equiv\frac{\sqrt3}{2}(\ket{L}-\ket{R})\big\}.
\eeq
The Hamiltonian, restricted to this basis and shifted by $-\alpha\eins/2$, reads
\beq
\hat H_{\rm bb}=\frac{\alpha}6\left(
\begin{array}{cc}
-1 & -2\sqrt2 x \\ -2\sqrt2 x & 1
\end{array}\right).\label{eqn:hbbmatrix}
\eeq
Corresponding eigenvectors are 
\begin{eqnarray}
\ket{-}_x&=&\cos\frac{\xi(x)}2\ket{b_1}+\sin\frac{\xi(x)}2\ket{b_2},\\
\ket{+}_x&=&\sin\frac{\xi(x)}2\ket{b_1}-\cos\frac{\xi(x)}2\ket{b_2},
\end{eqnarray}
with $\tan\xi(x)=2\sqrt2 x$. The energy eigenvalues are $\epsilon_{\mp}=\mp\alpha\sqrt{1+8x^2}/6$.
We have $\ket{-}_{x\rightarrow1}=\ket{L}$ 
and $\ket{-}_{x\rightarrow-1}=\ket{R}$. 
Since due to the adiabatic theorem the system remains in the same band of instantaneous eigenstates when
the parameter $x$ is changed slowly from $+1$ to $-1$, population can be transferred from $\ket{L}$ to $\ket{R}$
through adiabatic passage.
Notice that the time dependence in $\ket{\pm}_x$ and $\epsilon_{\pm}$ is
explicit through the control parameter $x(t)$.

\begin{figure}[t]
\includegraphics[width=0.7\linewidth,clip]{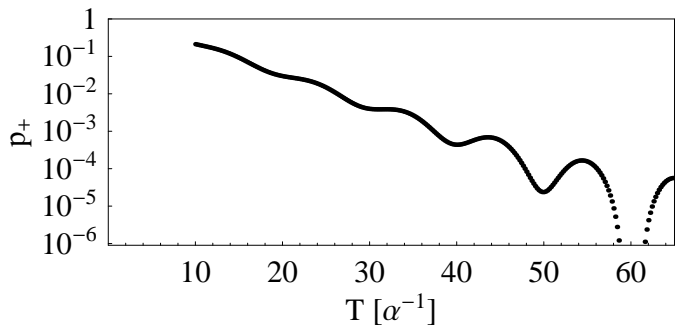}
\caption{
Excited state population $p_+$ obtained from Eq.~(\ref{eq:pplus}) as a function of the duration $T$ of the process.
}
\label{fig:pplus_3sites}
\end{figure}

A characterization of the adiabatic regime necessary to obtain transport with high fidelity
is provided by the condition \cite{guerin:2002}
\beq
T\gg \frac{\max_x|c(x)|}{\min_x\Delta(x)}.
\label{adiabatic}
\eeq
Here it is assumed that $x$ is changed at a constant rate, i.e., $dx/dt=2T^{-1}$, and
$T$ then is the total duration of the process. Furthermore,
\beq
c(x)\equiv{}_x\bra{+}\frac{d}{dx}\ket{-}_x=\frac{\sqrt2}{1+8x^2}\label{eq:c3sites}
\eeq
is the coupling between $\ket{-}_x$ and $\ket{+}_x$, and
\beq
\Delta(x)\equiv\epsilon_+-\epsilon_-=\frac{\alpha}3\sqrt{1+8x^2}\label{eq:delta3sites}
\eeq
is the energy difference of the two levels. Using Eqns. (\ref{eq:c3sites}) and (\ref{eq:delta3sites}) we obtain $T\gg3\sqrt2/\alpha$. In this limit,
first order perturbation theory allows to obtain an estimation of the final population $p_+$ of $\ket{+}$
due to the non-adiabatic coupling \cite{haensel:2001} ($\hbar=1$):
\beq
p_+(T)=\left|\int_{x=1}^{-1}dx\,\exp\left(i\,T\int_{x'=1}^xdx'\,
\Delta(x')\right)c(x)\right|^2.\label{eq:pplus}
\eeq
The excited state population $p_+$ is plotted in Fig.~\ref{fig:pplus_3sites} as a function of the total process
duration $T$. For $T$ large, the term $\exp(T\int_{x'=1}^xdx'\,\Delta(x'))$ oscillates rapidly, and due
to destructive interference the excited state population nearly averages to zero after one cycle. For this
reason, $p_+$ generically decreases as $T$ is increased. The additional dips in the curve correspond
to values of $T$ for which interference is such that $p_+$ is reduced after the whole process.

For $\ket{\phi}=\ket{0},\,\ket{-1}$ the same effective Hamiltonian (Eq.~\ref{eqn:hbbmatrix}; in the respective basis)
is obtained. Thus any state is transferred with the same efficiency and the same dynamical phase as $\ket{1}$.
This follows from the rotational symmetry of the Hamiltonian as well as of the initial state of the chain.
Additional terms in the Hamiltonian breaking the rotational symmetry will in general introduce differences in the dynamical
phases and a change in the adiabaticity condition. As an example, let us discuss the presence of a global magnetic field term
$\alpha h\sum_i\hat J^z_i$ in the Hamiltonian (\ref{eq:H2}). We assume $h$ to be small in order to still have a dimerized
ground state. The energy of the states $\ket{\pm}_x$ then is shifted by $\alpha hS_{\rm tot}$, where $-1\leq S_{\rm tot}\leq1$ is the 
total magnetization of the chain. Both, $\ket{+}_x$ and $\ket{-}_x$, are shifted by the same amount, such that the couplings
and thus the adiabaticity condition are not altered. However, the shift depends on the initial state $\ket{\phi}$, and a differential
dynamical phase is introduced. This can be corrected after the transfer by applying a local unitary to the
last spin. Dynamical phases from fluctuations of $h$ around a constant value nearly average out,
as long as they happen on a time scale short compared to the total process duration. Large field fluctuations on short
time scales additionally might spoil the adiabaticity, leading to larger couplings to excited states. 

We can immediately generalize this scheme for a larger chain with an odd number of spins.
For $x=1$, only even couplings are turned on. Then the ground state consists of a decoupled spin
at site $i=1$, and all other spins are paired into singlets:
\begin{equation}
	\ket{\psi_{x=1}^0(\phi)}=\ket{\phi}\ket{s}_{2,3}\ldots\ket{s}_{2i,2i+1}\ldots\ket{s}_{2n,2n+1}.
	\label{eq:psistart}
\end{equation}
On the other hand, if $x=-1$, such that only odd coupling are active,
then the ground state has again $n$ pairs of singlets, now with the last spin decoupled:
\begin{equation}
	\ket{\psi_{x=-1}^0(\phi')}=\ket{s}_{1,2}\ldots\ket{s}_{2i-1,2i}\ldots\ket{s}_{2n-1,2n}\ket{\phi'}.
	\label{eq:psiend}
\end{equation}
The magnetization $\vecb{M}=\ew{\sum_i\vecb{J}_i}$ of
$\ket{\psi_{x=-1}^0}$ and $\ket{\psi_{x=+1}^0}$ is completely fixed by the first
and the last spin, respectively, as the rest of the chain is non-magnetized.
Also, the magnetization is conserved under the rotationally
invariant set of Hamiltonians $\hat H_{\rm bb}(x)$, and moreover it can be shown that for any
$-1\leq x\leq 1$ the ground state is unique for a given magnetization $M_z$ \cite{yip:2003}. In a finite system,
it is furthermore separated by a gap from the first excited state. We can thus apply
the same protocol as for $N=3$ for transporting an unknown state of the first spin to the last one.
\begin{figure}[b]
\includegraphics[width=0.7\linewidth,clip]{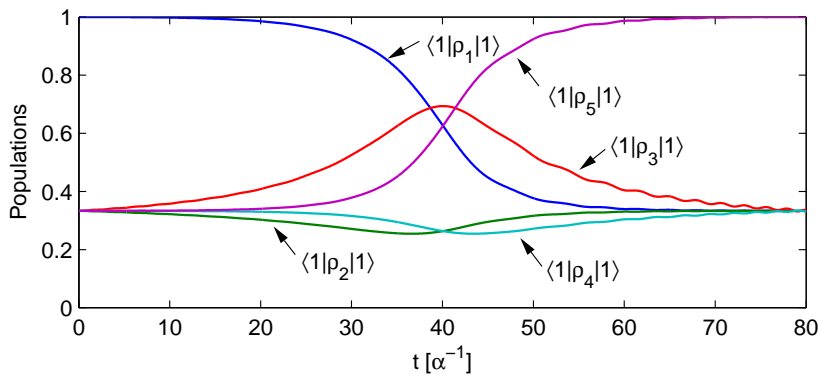}
\caption{
The populations $p_i(t)={}_i\langle1|\rho_i(t)|1\rangle_i$ for a chain of $5$ sites as a function of time, changing $x$
from $1$ to $-1$ at constant rate.
}
\label{fig:5sites_t80}
\end{figure}
An example of the population transfer via adiabatic passage is depicted in \figref{fig:5sites_t80} for a chain
of 5 sites. The first spin is initialized to $\ket{\phi}=\ket{1}$. Populations of $\ket{1}$ for site $i$,
$p_i(t)={}_i\langle1|\rho_i|1\rangle_i$  ($\rho_i$ is the reduced density matrix of site $i$), are shown as
a function of time as $x$ is changed linearly from $1$ to $-1$ in a time interval $T=80\alpha^{-1}$ ($\hbar=1$).
As expected, a significant increase of the population of $\ket{1}$ is only present for odd sites.
For perfect adiabatic evolution, we should find $p_{i}(t)=p_{N-i}(T-t)$.
The deviations (wiggles for $t>T/2$) stem from non-adiabatic transitions.

Let us discuss the conditions required to be in the adiabatic regime 
(Eq. (\ref{adiabatic}))  for a chain of length $N$.
Using $|\bra{\psi^1(x)}d/dx\ket{\psi^0(x)}|=|\bra{\psi^1(x)}(d\hat H_{\rm bb}/dx)\ket{\psi^0(x)}/\Delta(x)|$ and
\beq
\frac{d\hat H_{\rm bb}}{dx}=\alpha\sum_{i}\frac{(-1)^i}{2}\hat H_{\rm bb}^{(i,i+1)}
%\left[\cos\theta\left(\vecb{J}_i\vecb{J}_{i+1}\right)+
%\sin\theta\left(\vecb{J}_i\vecb{J}_{i+1}\right)^2\right]
	\label{eq:diffHbb}
\eeq
together with $||\hat P_0^{i,i+1}||=1$, $c(x)$ can be bound from above as $c(x)\leq(N-1)/\Delta(x)$.
The minimal gap $\min_x\Delta(x)$ is more difficult to estimate. For $N$ odd and $\theta=-\pi/2$
it is know from the equivalence of $\hat H_{\rm bb}(x=0)$ to an XXZ model, that in the 
thermodynamic limit, \ie, for $N\rightarrow\infty$, the gap vanishes:
$\lim_{N\rightarrow\infty}\Delta_{\rm min}=0$ \cite{albertini:2000}. For small chains, we find that the $\min_x\Delta(x)$ decreases
approximately linear in $1/N$. Then the minimal value $T$ necessary for being in the adiabatic regime grows approximately with the third power
of the number of sites.

\begin{figure}[t]
\includegraphics[width=0.7\linewidth,clip]{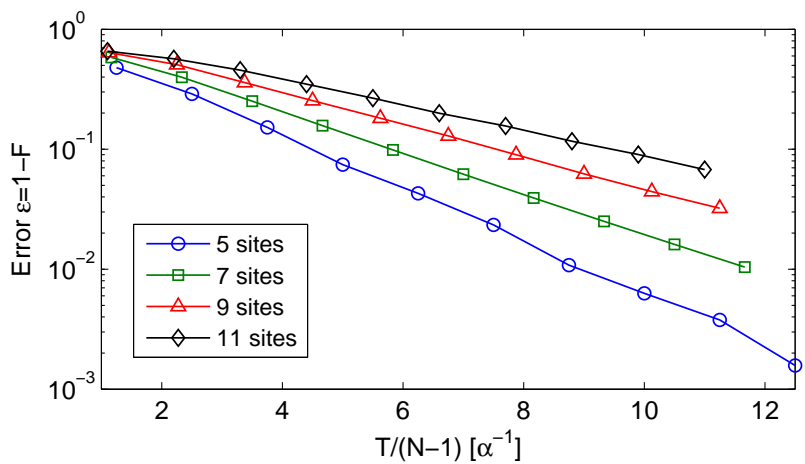}
\caption{
Logarithmic plot of the error $\epsilon=1-F=1-{}_N\bra{1}\rho_N(t)\ket{1}_N$ versus velocity $T/(N-1)$
for various lengths of the chain ($\theta=-\pi/2$). The couplings are changed linearly: $x(t)=(T-2t)/T$.
Time evolution has been performed using matrix product state algorithms \cite{garcia-ripoll:2006}.
}
\label{fig:fidvsNlin}
\end{figure}

The error of the transfer process is given by $\epsilon=1-F$, with $F={}_N\bra{1}\rho_N(T)\ket{1}_N$ being the transfer fidelity.
In \figref{fig:fidvsNlin}, the error $\epsilon$ is plotted for different durations of the process and for chains
of various lengths. Clearly, $\epsilon$ decreases as the transfer process is made slower.
The maximal velocity $T/(N-1)$ permitted for state transfer with fixed maximal error
decreases as $N$ is increased.

%Let us note a few features of this transport scheme using time-dependent couplings:
%(i) the fidelity can be made arbitrarily close to one by increasing the process duration $T$;
%(ii) if for $t>T_{\rm final}$ the coupling is left unchanged, then the state of the chain does
%not evolve in time, and especially no error can occur from inappropriate timing in decoupling the last
%spin from the chain as it is the case for schemes using static couplings of neighboring spins
%\cite{bose:2003:b,christandl:2004,subr:2004};
%(iii) as long as the change of the couplings is adiabatic, the transport fidelity does not
%depend on the exact path $x(t)$ in parameter space, nor on the exact choice of the Hamiltonian.
%To illustrate the latter feature, in Fig. xxx we show the transport fidelity
%$\bra{\phi}\hat\rho_{N}\ket{\phi}$, where
%$\hat\rho_{N}=\tr_{1,\ldots,N-1}\left(\ket{\psi(t_{\rm final})}\bra{\psi(t_{\rm final})}\right)$
%is the reduced density matrix of the last spin after the process.

\section{Optimization of the transfer fidelity}\label{sec:fid}

\begin{figure}[b]
\includegraphics[width=0.7\linewidth,clip]{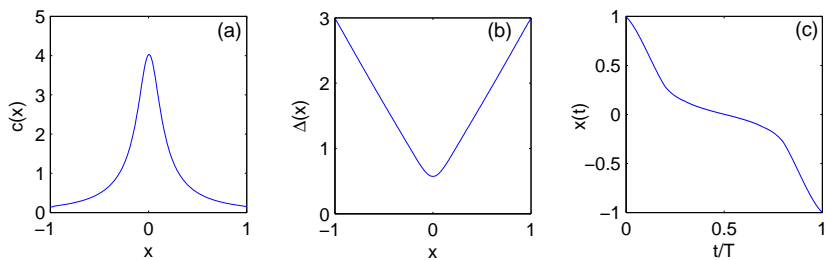}
\caption{
(a) Coupling $c(x)$ and (b) gap $\Delta(x)$ to the first
excited state as a function of the parameter $x$ for $\theta=-\pi/2$ and $N=7$ sites.
(c) The optimized path $x=x_{\rm opt}(t)$, obtained from the curves of (a,b).
}
\label{fig:DeltaxCx}
\end{figure}

For the Hamiltonian considered here,
the decrease of the transfer velocity as the number of sites is increased cannot be avoided,
since the gap to the first excited state closes for $N\rightarrow\infty$.
For small chains ($N\leq11$) we have performed an exact diagonalization of $\hat H_{\rm bb}(x)$
to obtain $c(x)$ and $\Delta(x)$. Figs.~\ref{fig:DeltaxCx} (a,b) show typical values
of $c(x)$ and $\Delta(x)$ for the coupling of the ground state to the first excited state (both with the same total spin).
The gap $\Delta(x)$ decreases approximately linear for $|x|$ varying from $1$ to $0$, while $c(x)$ is well
described by a Lorentzian function. Clearly, both are sharply peaked at $x=0$.
This suggest to use optimal control techniques to increase the (mean) transfer velocity
by adapting the change of the spin-spin couplings, \ie, $dx(t)/dt$, to the instantaneous energy difference $\Delta(x)$ and
the coupling $c(x)$.
In \cite{haensel:2001}, a method has been developed to obtain
an optimized path $x=x_{\rm opt}(t)$ which allows to suppress excitations to other
states, i.e., to reduce the transition probability 
\beq
p(T)=\left|\int_{t=0}^Tdt\exp\left(i\int_{t'=0}^tdt'\Delta(x(t'))\right)c(x(t))\frac{dx(t)}{dt}\right|^2.
\eeq
To this aim, we replace $\int_{t'=0}^tdt'\Delta(x'(t))=\frac{T}{T_0}\tau(t)$ (where $T_0$ is chosen
to have $\tau(0)=0$ and $\tau(T)=1$), such that the integral
becomes the Fourier transform of a function $u(\tau)\equiv\frac{dx(\tau)}{d\tau}c(x(\tau))$.
Now, for $u(\tau)$ we choose a Blackman pulse \cite{blackman} to reduce the side lobes of the Fourier transform.
The shape $x(t)$ is then obtained from solving two differential equations in order to obtain first
$x(\tau)$ (from inverting the equation for $u(\tau)$) and subsequently $\tau(t)$.

\begin{figure}[t]
\includegraphics[width=0.7\linewidth]{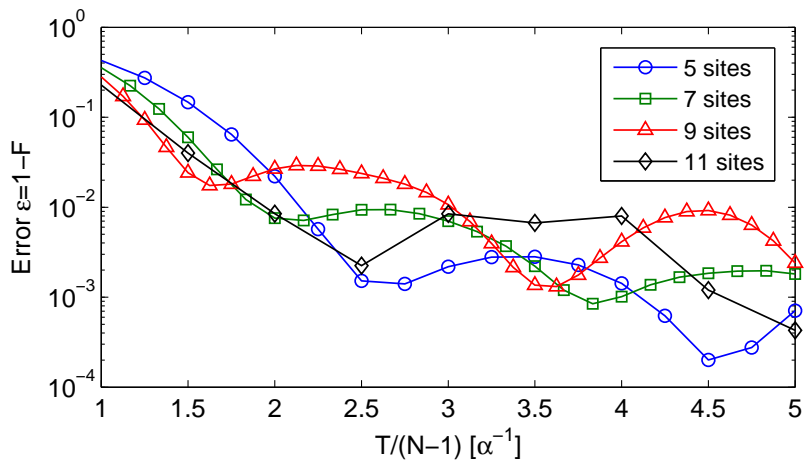}
\caption{
Error $\epsilon=1-F$ versus (mean) velocity $T/(N-1)$ for various lengths of the chain,
$\theta=-\pi/2$. An optimized path $x=x_{\rm opt}(t)$ (see Fig. \ref{fig:DeltaxCx}) is used to changed the coupling in time.
}
\label{fig:fidvsNopt}
\end{figure}

An example of a path $x(t)$ obtained in such a way is plotted in \figref{fig:DeltaxCx} (c).
The dependence of the transfer error on
the mean velocity $T/(N-1)$ if such an optimized path is used, is plotted in \figref{fig:fidvsNopt}.
Compared to the results for a linear change of $x$ in time, displayed in Fig.~\ref{fig:fidvsNlin}, the error
for a given mean velocity is typically more than one magnitude smaller for the optimized path.
%We have used a linear fit in the finite-size analysis of $\Delta$ as well as of the parameters of the Lorentzian
%fitted to $c(x)$, to calculate optimized paths $x_{\rm opt}(t)$ also for larger $N$.

\section{Experimental implementation}\label{sec:impl}

Let us finally discuss the implementation of the bilinear--biquadratic Heisenberg Hamiltonian and the realization
of the spatial--temporal control of the couplings.
We consider atoms with spin $F=1$ (\ie, atoms with an $2F+1$-dimensional hyperfine degree
of freedom) at low temperatures confined in a deep (1D) optical lattice. The system then is
well described \via the Bose-Hubbard Hamiltonian \cite{imambekov:2003,lewenstein:2006}
\begin{equation}
\hat H_{\rm BH}=-t\sum_{\ew{ij},\sigma}\left[\hat a_{i,\sigma}^{\dagger}\hat a_{j,\sigma}+{\rm h.c.}\right]
+\frac{c_0}2\sum_i\hat n_i(\hat n_i-1)+\frac{c_2}2\sum_i (\vecb{F}_i^2-2\hat n_i).
\end{equation}
Here $\hat a_{i,\sigma}$ annihilates a particle in a hyperfine state with $m_F=\sigma$ at site $i$,
$\hat n_i=\sum_{\sigma}\hat a_{i,\sigma}^{\dagger}\hat a_{i,\sigma}$ is the number of particles, and
$\vecb{F}_{i}=\sum_{\sigma\sigma'}\hat a^{\dagger}_{i,\sigma}\vecb{T}_{\sigma\sigma'}\hat a_{i,\sigma'}$
is the (total) spin operator on site $i$ ($\vecb{T}$ being the usual spin-$1$ matrices). The tunneling
amplitude $t$ is obtained from the overlap of the Wannier functions $w(x)$ \cite{lewenstein:2006}. Tunneling
conserves both, the total spin $S$ and the total spin projection $m_S$. The parameters $c_0=(g_0+2g_2)/3$
and $c_2=(g_2-g_0)/3$ depend on the effective 1D interaction strengths in the spin $S$ channel, $g_S$. Those
are proportional to the two-body spin-dependent scattering length $a_S$. Their absolute strength can be controlled
\via the orthogonal confinement, the relative strength moreover can be changed \via (magnetic or optical)
Feshbach resonances. Note that due to the bosonic character of the particles, on-site contact interactions
are zero for odd total spin.
%\beq
%g_S=\frac{4\pi\hbar^2a_S}M\left[\int dy |w(y)|^4\right]^2,
%\eeq
%with $a_S$ being the scattering length and $M$ the atomic mass. $w{\perp}(y)$ is the Wannier function
%for the $y,\,z$--directions. 

The ratios $t/c_0,\,t/c_2$ between tunneling and on-site interactions
are tunable \via the lattice parameters (lattice depth and orthogonal confinement). For $t\ll |c_0|,\,|c_2|$ the system is 
in a Mott-insulating state with atoms being quenched at fixed lattice sites. Here we will only consider the 
case of having a single particle per lattice site \cite{rizzi:2005}, and assume tunneling to be sufficiently weak
compared to on-site interactions, such that it can be treated perturbatively with $t/c_0$ as small parameter.
Then, the following effective spin-spin Hamiltonian can be obtained in second order perturbation theory \cite{imambekov:2003}:
%considering the two-site case, states with total spin $S$ in second order perturbation theory obtain a phase shift
%$\epsilon_S=-4t^2/g_S$, such that the resulting effective spin-spin Hamiltonian in second order reads
\begin{equation}
\hat H_{\rm eff}^{i,i+1}=-\frac{2t^2}{c_0}\left[\frac1{1+\tilde c_2}\left(\vecb{J}_i\vecb{J}_{i+1}\right)
+\frac13\left(\frac1{1+\tilde c_2}+\frac2{1-2\tilde c_2}\right)\left(\vecb{J}_i\vecb{J}_{i+1}\right)^2\right]\label{eq:Heff},
\end{equation}
where $\tilde c_2=c_2/c_0$.
This Hamiltonian is equivalent to the one of Eq.~(\ref{eq:H1}). For ${}^{23}$Na,
the bare values of the scattering lengths are very similar: $a_0=46a_B,\,a_2=52 a_B$ \cite{burke:1998}. Then $\tilde c_2\approx0.04$ and
$\theta\approx-0.74\pi$. However, through auxiliary magnetic 
and electric fields, it is possible to modify the system such that
the complete anti-ferromagnetic part of the phase diagram can be reached \cite{garcia-ripoll:2004}.

\begin{figure}[b]
\includegraphics[width=0.6\linewidth]{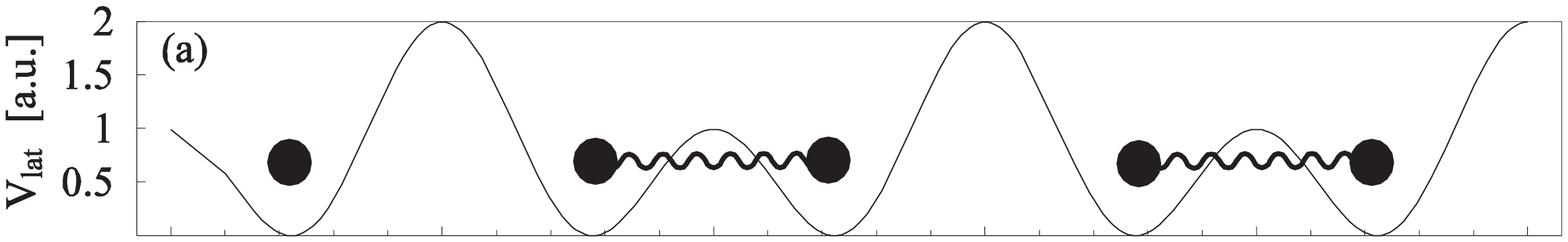}\\
\includegraphics[width=0.6\linewidth]{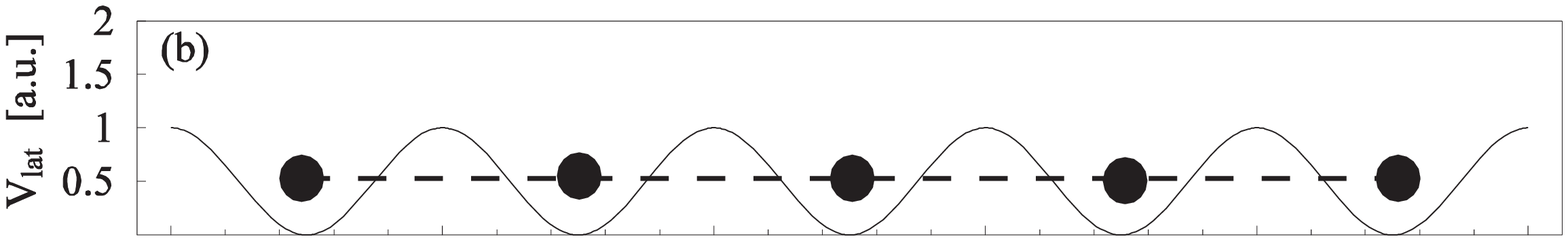}\\
\includegraphics[width=0.6\linewidth]{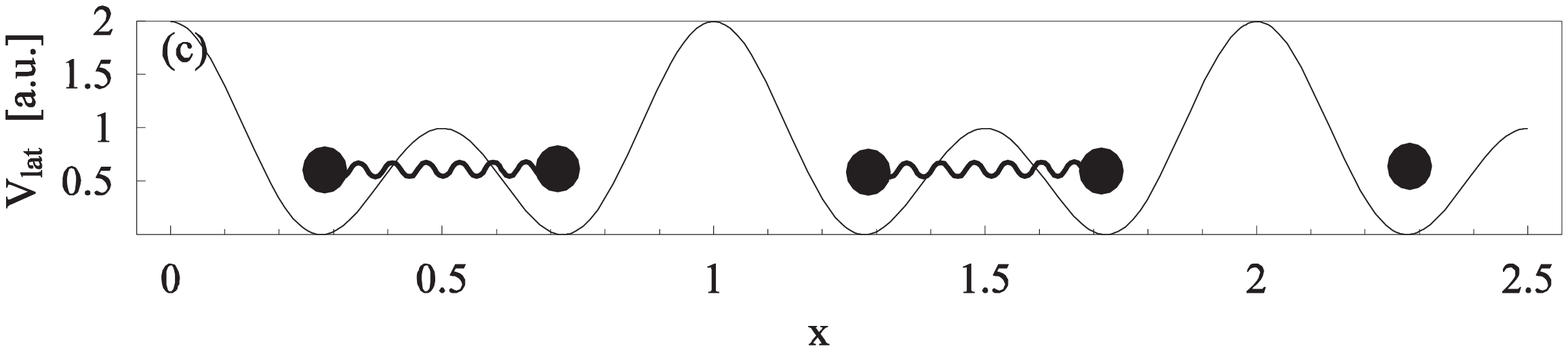}
\caption{
Sketch of the optical superlattice potentials obtained
from varying the laser intensities $I_{\rm half}$ and $I_{\rm full}$
to obtain the required control over the couplings.
(a) $I_{\rm full}>0,\,I_{\rm half}>0$, $\phi_{\rm full}=\pi/2$; 
(b) $I_{\rm full}=0,\,I_{\rm half}>0$, $\phi_{\rm full}=\pi/2$; 
(c) $I_{\rm full}>0,\,I_{\rm half}>0$, $\phi_{\rm full}=0$ (see text for details). 
}
\label{fig:vlat}
\end{figure}

Let us now move to the realization of the adiabatic passage.
The necessary spatial-temporal variation of the couplings can be realized by two pairs of
laser beams of wavelengths $\lambda$ and $\lambda/2$, respectively, and identical polarizations, counter-propagating in $x$ direction
(additional optical potentials in $y$- and $z$-direction are necessary to confine the atoms in 3D). The trapping potential in $x$-direction seen by the atoms then is proportional to the square of the electric field. Thus
\begin{equation}
V_{\rm lat}(x,t)\,\propto\,I_{\rm half}(t)\cos^2(2\pi x /(\lambda/2))+
I_{\rm full}(t)\cos^2(2\pi x/\lambda+\phi_{\rm full}(t)),
\end{equation}
where $I_{\rm half}$ and $I_{\rm full}$ are the corresponding laser intensities.
We have introduced an additional phase shift $\phi_{\rm full}$ for the laser of wavelength $\lambda$.
The effective spin-spin-coupling is proportional to $t^2$, see Eq.~(\ref{eq:Heff}), which
in turn has an exponential dependence on the height of the potential between adjacent sites. Setting only
$I_{\rm half}>0$ (and $I_{\rm full}=0$), a lattice (with a distance of $\lambda/4$ between
adjacent sites) is defined with all couplings being equal. We assume this lattice to be loaded with
a single particle per site in the ground state. Setting $\phi_{\rm full}=\pi/2$, then $I_{\rm full}$ is increased
to strongly reduce the coupling between each second pair of sites. At the same time, $I_{\rm half}$ has to be adjusted to
keep the tunneling between the other sites in a regime where the effective Hamiltonian (\ref{eq:Heff}) is valid.
This provides the initial situation for the transport process, {\it c.f.}, Fig.~\ref{fig:vlat} (a). After preparation of the first spin,
$I_{\rm full}$ is decreased to zero (adjusting properly also $I_{\rm half}$; see Fig.~\ref{fig:vlat} (b)), and $\phi_{\rm full}$ is set to zero.
Then increasing $I_{\rm full}$ again allows to turn off selectively only the other subset of spin-spin-couplings [Fig.~\ref{fig:vlat} (c)].
This procedure realizes the change 
of the Hamiltonian necessary for the adiabatic passage transfer process.

Finally, we can make a rough estimation of the time scales of the process which can be achieved with realistic optical lattice parameters.
For ${}^{23}$Na, $\lambda/2=514\,$nm, and for tight orthogonal confinement, in the perturbative
regime ($|t/c_0|\ll1$) couplings $\alpha=\hbar J$ with $J\approx100\ldots200\,$s$^{-1}$ could be achieved,
leading to $T_{N=9}\approx(1\ldots25)\cdot10^{-2}\,$s
for $N=9$ sites and an error $\epsilon<10^{-2}$ in the optimized case. 
For atoms with spin $F=1$, like Sodium, deep in the Mott insulating phase,
decoherence is caused either by 3-body losses or by scattering of photons.
Notice, however, that 3-body losses are relevant only for 
filling factors larger than two.
If the lasers forming the optical lattice are far detuned from all atomic
resonances, the spontaneous scattering rate is highly suppressed
and coherence times much larger than $T_{N=9}$ are easily accessible.
That means that the transport scheme we propose via adiabatic passage
in spin chains should be experimentally realizable.

\section{Conclusions}\label{sec:concl}

In this article we have presented a scheme to transfer a quantum state through a
spin chain using an adiabatic passage technique. The most remarkable features of
this scheme are the following. First, the transfer fidelity can be made arbitrarily
close to one by increasing the transfer time. Second, as long
as the change of the couplings is adiabatic, the transport fidelity does not
depend on the exact path $x(t)$ in parameter space. There is thus no need of a precise control
of couplings and timing. Third, in contrast to other transfer schemes in spin chains, there is no need for a receiver,
meaning that once the transfer has been accomplished the system remains frozen. 

We have applied this method to a spin-$1$ chain in the anti-ferromagnetic dimerized phase.
Using standard optimal control techniques, we have shown that the transfer velocity can be
substantially increased if an optimized path in parameter space is chosen.
We have proposed a realistic experimental implementation of adiabatic passage transfer
using ultracold atoms and optical superlattices.

In the spin-$1$ dimerized phase, the transfer velocity is limited by the fact that for a chain with an odd number of sites
the gap vanishes as $N\rightarrow\infty$. Identifying a system with similar characteristics but
with a gap that persists in the thermodynamic limit would allow
for finite transfer velocities regardless the size of the chain.

\section*{Acknowledgments}

We thank  Diego Porras for enlightening discussions.
We acknowledge support from MEC (Spanish Government) under contracts  AP2005-0595, FIS 2005-04627, -014697, -01369, EX2005-0830, CIRIT SGR-00185, CONSOLIDER-INGENIO2010  CSD2006-00019 ``QOIT".

%\bibliographystyle{unsrt_dis}
%\bibliography{ref}

\end{document}